\documentclass{goose-article}

\hypersetup{pdfauthor={%
  T.W.J. de Geus%
}}
\header{%
  T.W.J.~de~Geus, R.H.J.~Peerlings, M.G.D.~Geers\\%
  Procedia Materials Science, 2014, 3:598-603, %
  \doi{10.1016/j.mspro.2014.06.099}%
}

\title{%
  Topology and morphology influences on the onset of ductile failure in a two-phase microstructure
}
\author[1,2]{T.W.J.~de~Geus$^*$}
\author[1]{R.H.J.~Peerlings}
\author[1]{M.G.D.~Geers}
\affil[1]{%
  Department of Mechanical Engineering, Eindhoven University of Technology, Eindhoven, The Netherlands%
}
\affil[2]{%
  Materials innovation institute (M2i), Delft, The Netherlands%
}
\contact{%
  $^*$Corresponding author:
  \href{mailto:t.w.j.d.geus@tue.nl}{t.w.j.d.geus@tue.nl} %
  \hspace{1mm}--\hspace{1mm} %
  \href{mailto:tom@geus.me}{tom@geus.me} %
  \hspace{1mm}--\hspace{1mm} %
  \href{http://www.geus.me}{www.geus.me}%
}

\begin{document}

\maketitle

\begin{abstract}
Multi-phase material are frequently applied in a wide variety of products, as they posses a unique set of properties by combining two or more distinct phases at the level of the microstructure. Although the macroscopic stiffness and hardening are reasonably well understood, questions remain about the dominant failure mechanism(s). We identify the role of the microstructural topology (the distribution of phases) on damage ``hot-spots'' in the microstructure, by performing a numerical study on a large set of randomly generated topologies. The result identifies a distinct probability distribution of phases around a typical damage ``hot-spot''. This work is focused on assessing the sensitivity of the result to the assumptions made on the microstructural geometry.
\end{abstract}


\section{Introduction}

Multi-phase materials, such as dual phase steel, metal matrix composites, etc., are frequently used in engineering applications. These materials often posses both strength and ductility, by combining two or more phases at the level of the microstructure, for example hard (but brittle) particles embedded in a soft (ductile) matrix. Although their macroscopic stiffness and hardening may be reasonably well predicted for a given microstructure, many uncertainties remain about the dominant failure mechanism(s).

Several author have studied these failure mechanisms, see \cite{Williams2010, Kumar2006, Ghadbeigi2010} and others. However, only very few studies have performed a systematic analysis of the effect of the microstructural topology of mutli-phase materials. \cite{Kumar2006} generate representative microstructure -- based on statistics -- and then try to identify a critical configuration for damage. In the present study, similar to \cite{DeGeus2015a}, we focus on characterizing the distribution of phases around the damage hot-spot more systematically, at the onset of ductile failure of the soft phase. We compare an ensemble of random topologies by characterizing the topological ``hot-spot'' responsible for high damage. The resulting probability distribution of hard phase around the damage ``hot-spot'' displays a high probability of hard phase in the principal strain direction and soft phase under $\pm 45$ degree angles to this direction. The former introduces a high level of hydrostatic stress and the latter promotes plastic deformation in shear bands through the soft phase.

The purpose of this contribution is to verify some of the assumptions we have made in \cite{DeGeus2015a}. In that work, use was made of idealized geometries whereby individual particles were modeled as square elements. This morphology may lead to phase particles connected by a singular point, which is clearly unphysical. Such connection cannot occur in hexagonal elements, the element morphology used in this paper. Furthermore, equi-sized hexagonal elements lead to non-square unit cells, allowing us to uncover possible artifacts caused by the employed periodic boundary conditions.

\section*{Nomenclature}

\begin{tabular}{ll}
$\bm{A}$
&
second order tensor $A_{ij}$
\vspace*{1mm} \\
$\vec{a}$
&
vector $a_i$
\vspace*{1mm} \\
$\lfloor \ldots \rfloor$
&
positive values: $\lfloor x \rfloor = \tfrac{1}{2} \big( \, x + |x| \, \big)$
\vspace*{1mm} \\
$\langle \ldots \rangle$
&
ensemble average
\end{tabular}

\section{Modeling}
\label{sec:modeling}

\begin{figure}[htp]
  \begin{minipage}[b]{.50\textwidth}
  	\includegraphics[width=1.\textwidth]{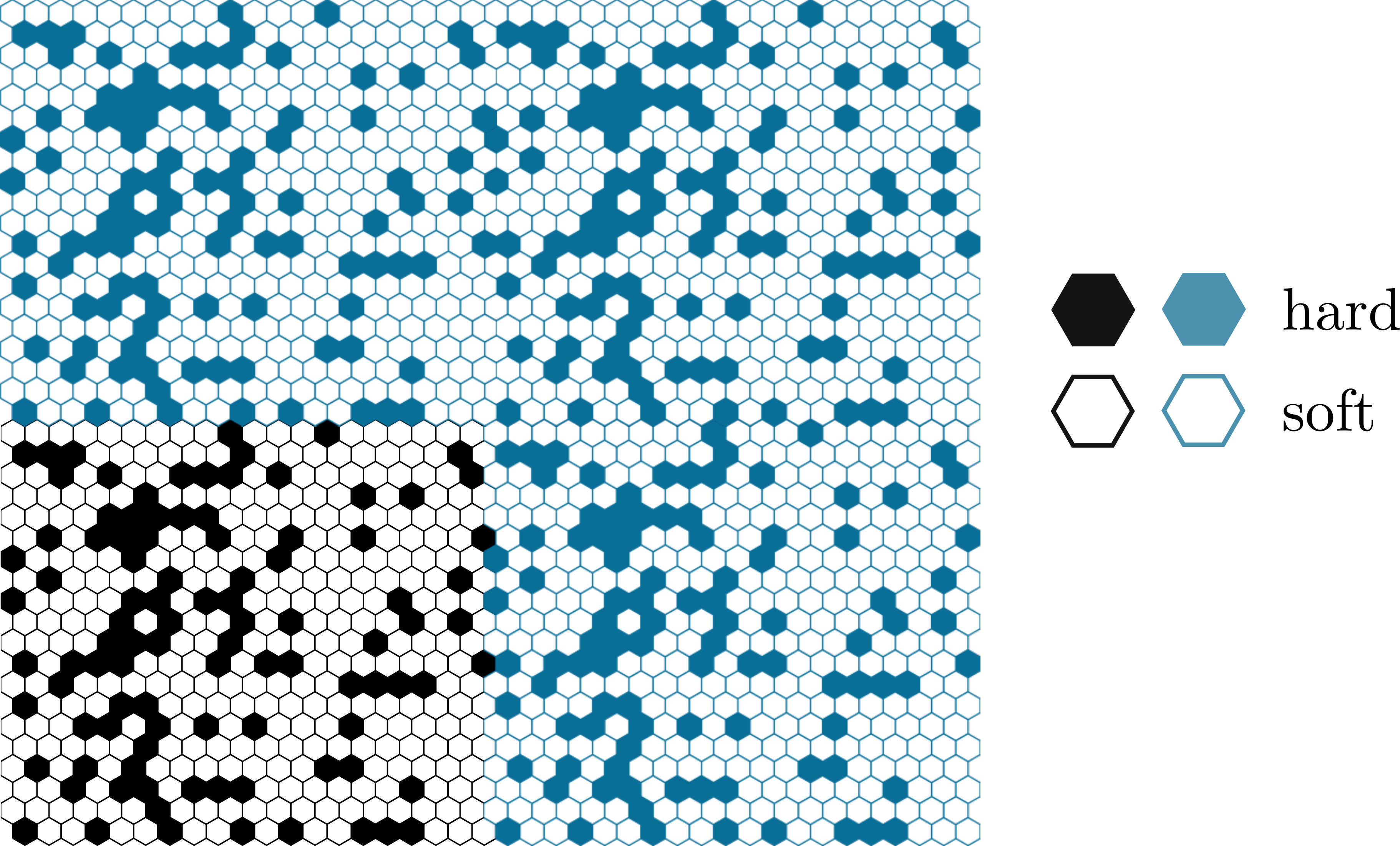}
  	\\ \footnotesize
  	(a) example cell, including periodicity
  \end{minipage}
  \hspace{.05\textwidth}
  \begin{minipage}[b]{.35\textwidth}
  	\centering
  	\includegraphics[width=.5\textwidth]{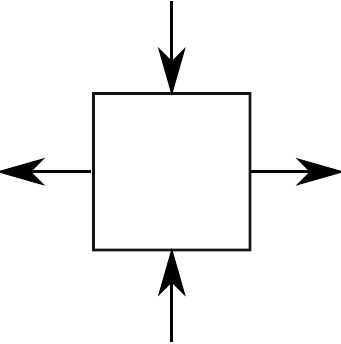}
  	\\ \footnotesize
  	(b) applied pure shear deformation
  \end{minipage}
  \caption{(a) Randomly generated unit cell of hexagonal elements. The cell is shown in the lower left corner in black (hard) and white (soft); the assumed periodicity is shown in blue/white. (b) The applied pure shear deformation.}
  \label{fig:geometry}
\end{figure}

The two considered phases are assumed to differ in their plastic response, whereby one is harder than the other. We consider an ensemble of $400$ different microstructures -- represented by unit cells. Each of these individual cells comprises $20 \times 20$ hexagonal elements to which we randomly assign the properties of the hard or the soft phase, according to a probability of $P^\text{hard} = 0.25$. We furthermore assume the unit cells to be periodic, whereby the periodicity is much larger than the size of the individual elements. An example of a cell is shown in Figure~\ref{fig:geometry}(a), where the cell is shown in black (hard) and white (soft), and its periodic extension in blue/white.

We assume both the hard and soft phase to be elasto-plastic. Since the local deformations may be significant, we consider a finite strain framework; in particular we use the model due to \cite{Simo1992a}. We furthermore assume isotropic $J_2$-plasticity with linear hardening, characterized by the initial yield stress $\tau_\text{y0}$ and the hardening modulus $H$, which are different for both phases.

We chose a set of material parameters representative for one type of multi-phase materials -- dual phase steel (e.g.\ \cite{Sun2009}). However, we consider them to be representative for a wider class of materials. The contrast between the phases is in terms of a difference in yield stress and hardening modulus as follows:
\begin{equation}
	\tau_\text{y0}^\text{soft} / E = 2 \cdot 10^{-3} \qquad H^\text{soft} / E = 5 \cdot 10^{-3}
	\qquad\qquad
	\tau_\text{y0}^\text{hard} / E = 6 \cdot 10^{-3} \qquad H^\text{hard} / E = 15 \cdot 10^{-3}
\end{equation}
$E$ is the Young's modulus and the Poisson's ratio $\nu = 0.3$. Note that the elastic constants are equal for both phases.

We restrict ourselves to damage initiation in the ductile, soft, phase. We choose a simple damage indicator which accounts for the influence of plastic strain and hydrostatic stress on ductile failure (cf.\ \cite{Rice1969, Gurson1977} and others). In this model the damage indicator $D$ is given by the product of the positive part of the hydrostatic stress $\tau_\text{m}$ and the effective plastic strain $\varepsilon_\text{p}$:
\begin{equation}
  D = \lfloor \tau_\text{m} \rfloor\, \varepsilon_\text{p}
\end{equation}
It is important to remark that this damage parameter does not affect the constitutive response, i.e.\ softening due to the damage is not included in the model.

The response of the individual cells is calculated using the finite element method. The implementation details for the particular constitutive model used here are found in \cite{Geers2004}; we however do not use the damage formulation here. Each hexagon is discretized by three eight-node quadratic finite elements. The response is then considered as the volume average of the integration points in these three finite elements, i.e.\ we consider only (microstructural) element average quantities.

\section{Periodicity conditions}
\label{sec:periodicity}

The assumed periodicity is reflected in the fact that periodic boundary conditions are employed. Nodes on one edge ($+$) are tied to nodes on the opposite edge ($-$) as follows
\begin{equation}
	\vec{u}^+ - \vec{u}^- = \big( \bar{\bm{F}} - \bm{I} \big) \cdot \big( \vec{X}_0^+ - \vec{X}_0^- \big)
\end{equation}
where $\vec{u}$ denotes the displacement of a node and $\vec{X}_0$ its position in the reference (undeformed) configuration. For the macroscopic deformation gradient $\bar{\bm{F}}$ we adopt pure shear, according to
\begin{equation}\label{eq:F}
	\bm{\bar{F}} = \exp \Big( \tfrac{\sqrt{3}}{2} \bar{\varepsilon} \Big)\,\, \vec{e}_x \vec{e}_x + \exp \Big( - \tfrac{\sqrt{3}}{2} \bar{\varepsilon} \Big)\,\, \vec{e}_y \vec{e}_y + \vec{e}_z \vec{e}_z
\end{equation}
where $\bar{\varepsilon}$ is the overall equivalent logarithmic strain (illustrated in Figure~\ref{fig:geometry}(b)). We incrementally apply a maximum deformation of $\bar{\varepsilon} = 0.10$.

\begin{figure}[htp]
  \centering
  \includegraphics[width=.46\textwidth]{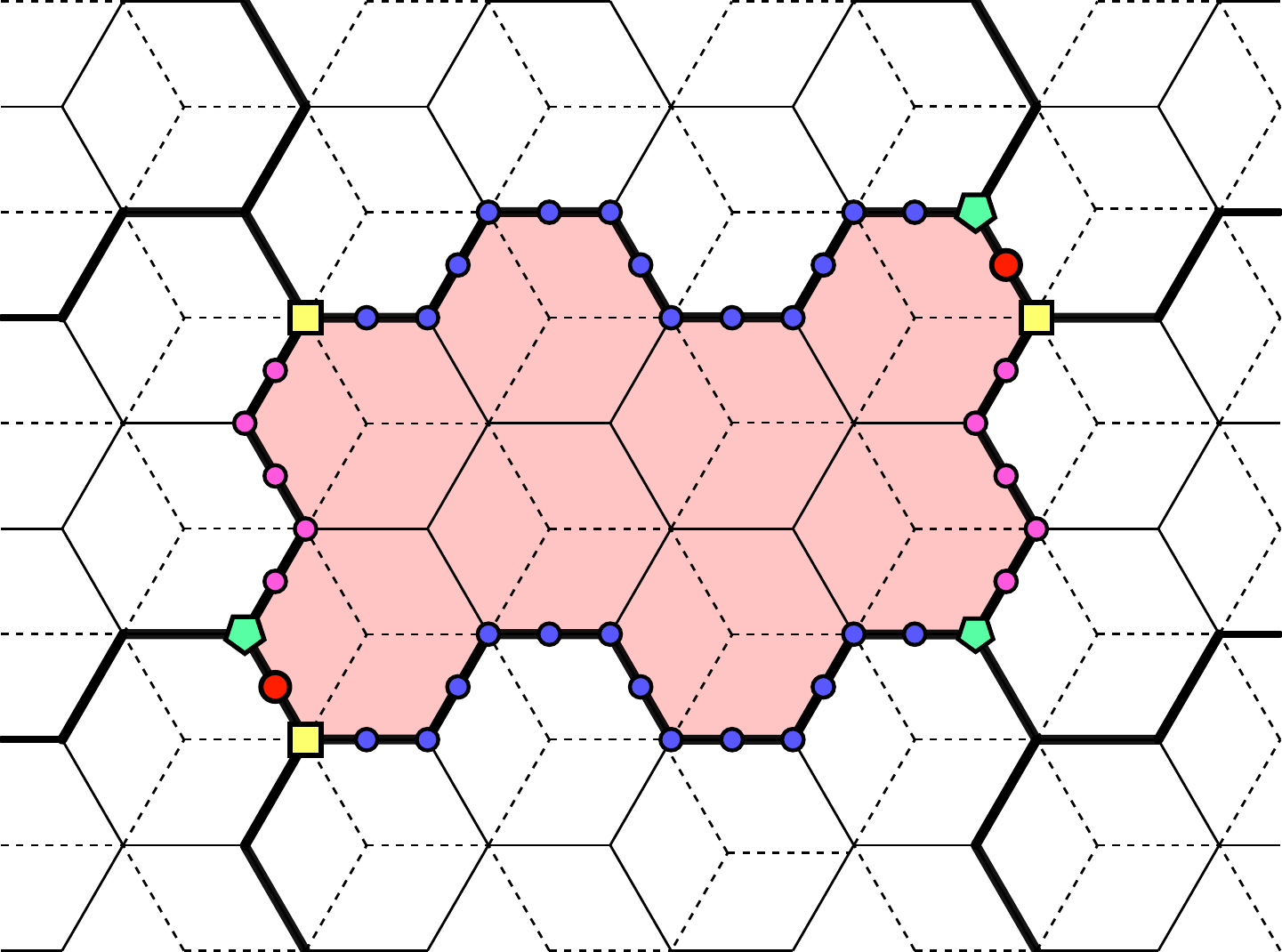}
  \caption{Spatial discretization (dashed lines) of the periodic cell -- highlighted in red -- comprising $4 \times 2$ hexagons (solid lines). The nodal tyings are shown using corresponding markers.}
  \label{fig:periodicity_hex}
\end{figure}

Some remarks are in order regarding the spatial discretization and the nodal tyings necessary to implement the periodicity condition for a cell with hexagonal elements. A simple illustration is given in Figure~\ref{fig:periodicity_hex}. In this figure the periodic cell is highlighted in red. The hexagonal elements are indicated by a solid black line and the finite element discretization by dashed lines. The nodal tyings are indicated using markers with identical shape and color. As is observed, the treatment of corners of the periodic cell is somewhat different than in the more common rectangular cells. In particular, in the lower-left and top-right part of the unit cell, two node combinations exist in which three nodes are coupled (for rectangular cells only one such combination exists). In between, one node is coupled between these two corners. The smaller red and blue nodes are coupled pairwise between left--right and top--bottom respectively.

\section{Simulation results}
\label{sec:results}

\begin{figure}[htp]
  \centering
  \begin{minipage}[b]{.40\textwidth}
  	\includegraphics[height=50mm]{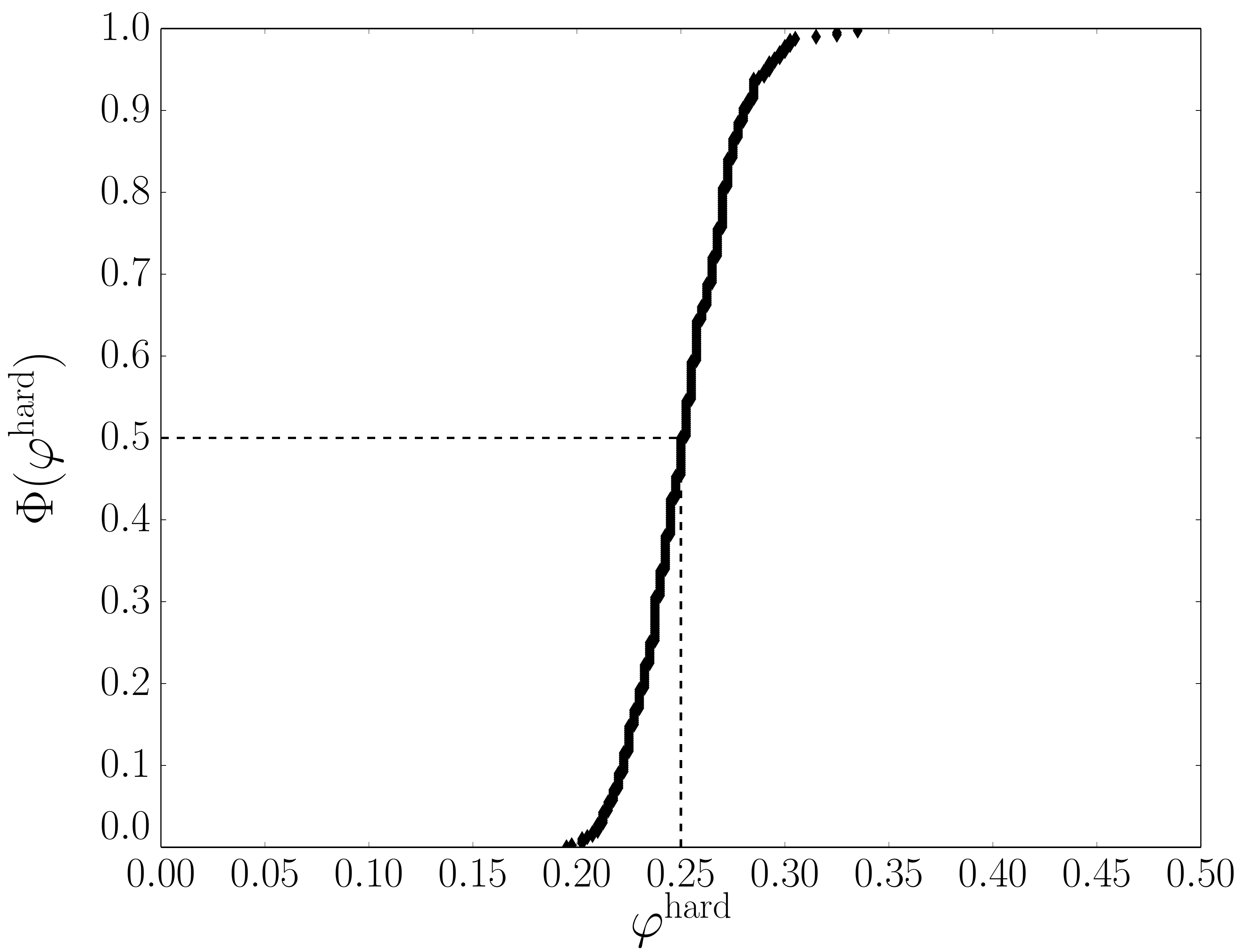}
  	\\ \footnotesize
  	(a) cumulative distribution of the volume fraction
  \end{minipage}
  \hfill
  \begin{minipage}[b]{.58\textwidth}
  	\begin{minipage}[c]{.69\textwidth}
  		\includegraphics[height=50mm]{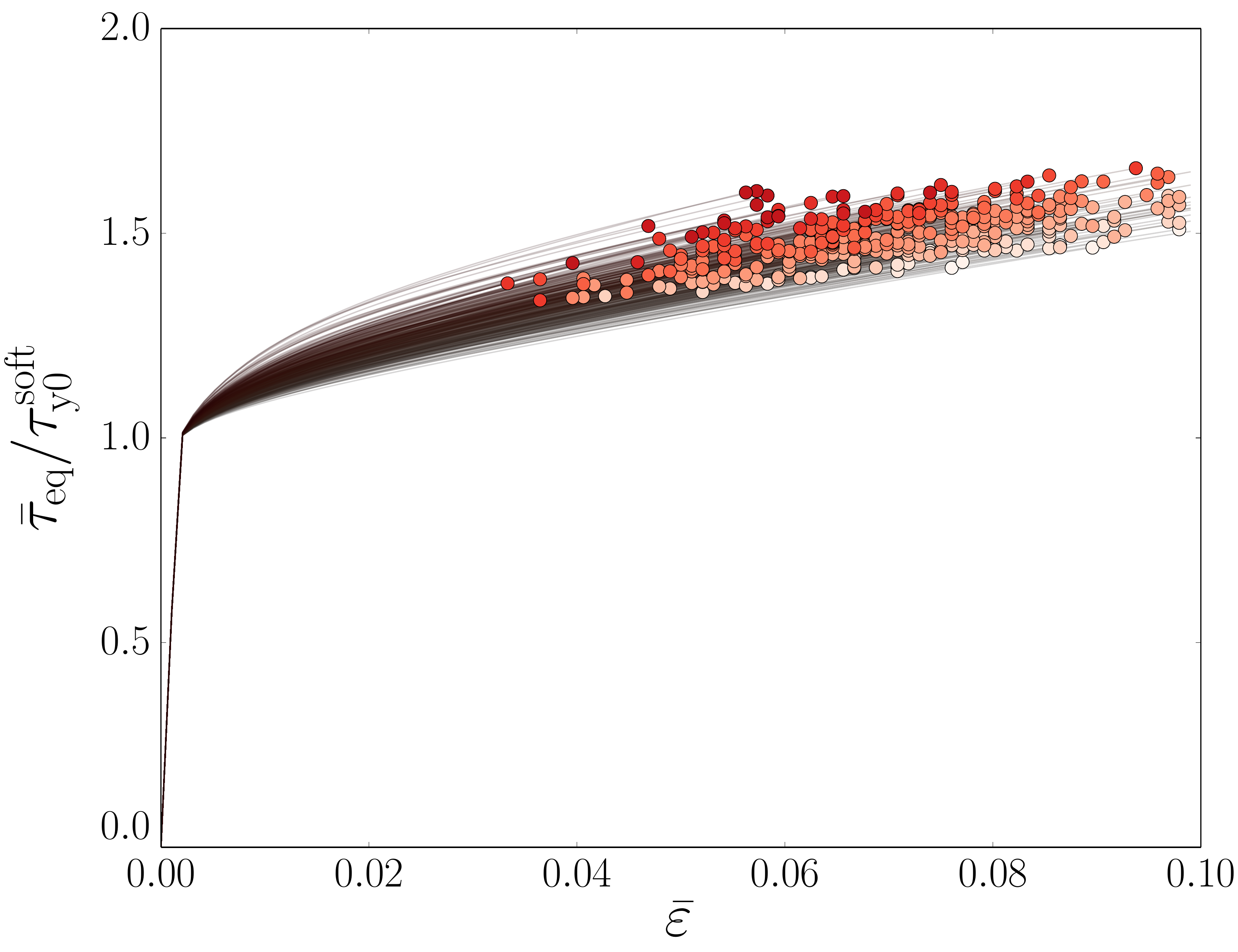}
    \end{minipage}
  	\begin{minipage}[c]{.29\textwidth}
  		\includegraphics[height=40mm]{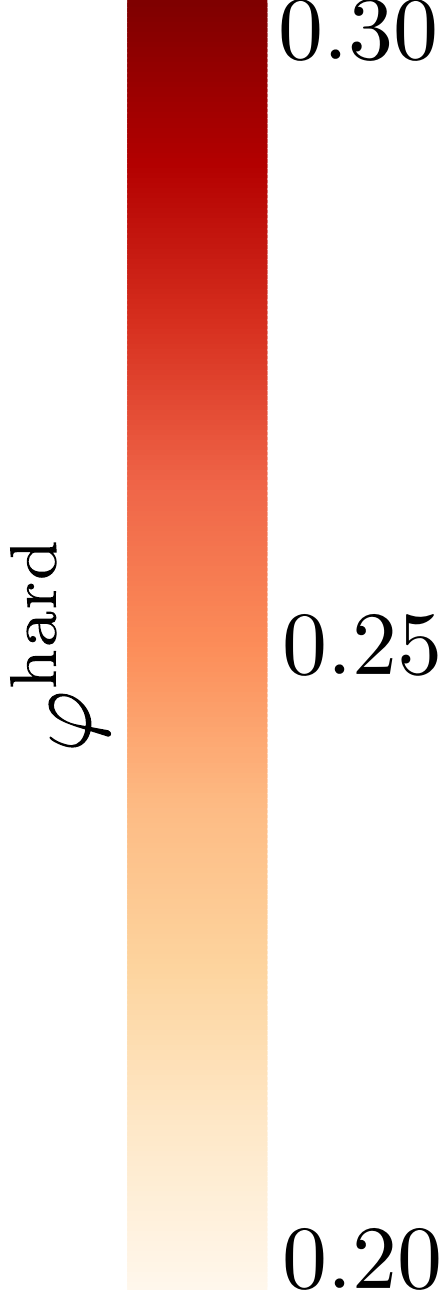}
  	\end{minipage}
  	\vspace*{1.0mm}
  	\\ \footnotesize
  	\hspace*{6mm} (b) macroscopic response, truncated at a critical damage level
  \end{minipage}
  \caption{(a) Cumulative distribution $\Phi$ of the hard phase volume fraction $\varphi^\text{hard}$ in the ensemble of $400$ random topologies. The ensemble average, $\langle \varphi^\text{hard} \rangle$, is indicated by dashed lines. (b) The macroscopic equivalent stress $\bar{\tau}_\text{eq}$ -- normalized by the initial yield stress of the soft phase $\tau_\text{y0}^\text{soft}$ -- of the individual cells as a function of the applied logarithmic strain $\bar{\varepsilon}$, truncated at the critical damage. The colors represented the hard phase volume fraction, $\varphi^\text{hard}$ (note: the extremes in volume fraction are outside the colorbar; to the unit cells that do not reach a critical damage level no marker is assigned).}
  \label{fig:volfrac}
\end{figure}

We begin our discussion of the results by considering the overall response of the individual cells. For the ensemble considered here, the variation in hard phase volume fraction is shown in Figure~\ref{fig:volfrac}(a) as a cumulative distribution. As observed, the hard phase volume fraction ranges between $\varphi^\text{hard} = 0.19$ and $0.34$ around an ensemble average of $\langle \varphi^\text{hard} \rangle \equiv P_\text{hard} = 0.25$.

The effect of the distribution in volume fraction is directly observed in the macroscopic responses of the cells. The macroscopic responses of the individual cells are shown in Figure~\ref{fig:volfrac}(b), where the normalized macroscopic equivalent stress $\bar{\tau}_\mathrm{eq}$ has been plotted as a function of the applied logarithmic strain $\bar{\varepsilon}$. The curves have been truncated at an, arbitrarily chosen, critical damage level of $D_\text{crit} = 100 \tau_\text{y0}^\text{soft} \varepsilon_\text{y0}^\text{soft}$. This point of ``fracture initiation'' is shown as a colored marker, whereby the color corresponds to the volume fraction of that particular unit cell.

When we consider this graph in more detail, we observe that all cells yield at the initial yield stress of the soft phase. The hardening -- a non-linear combination between the hardening of the two phases -- furthermore is stronger for a higher volume fraction. When we consider the strain at which a critical damage level is reached, $\bar{\varepsilon}_\text{crit}$, we observe a large scatter. From the colors we observe that $\bar{\varepsilon}_\text{crit}$ is largely uncorrelated to the volume fraction. It is indeed the difference in topology which is responsible for the large scatter in $\bar{\varepsilon}_\text{crit}$.

\begin{figure}[htp]
  \centering
  \includegraphics[width=1.\textwidth]{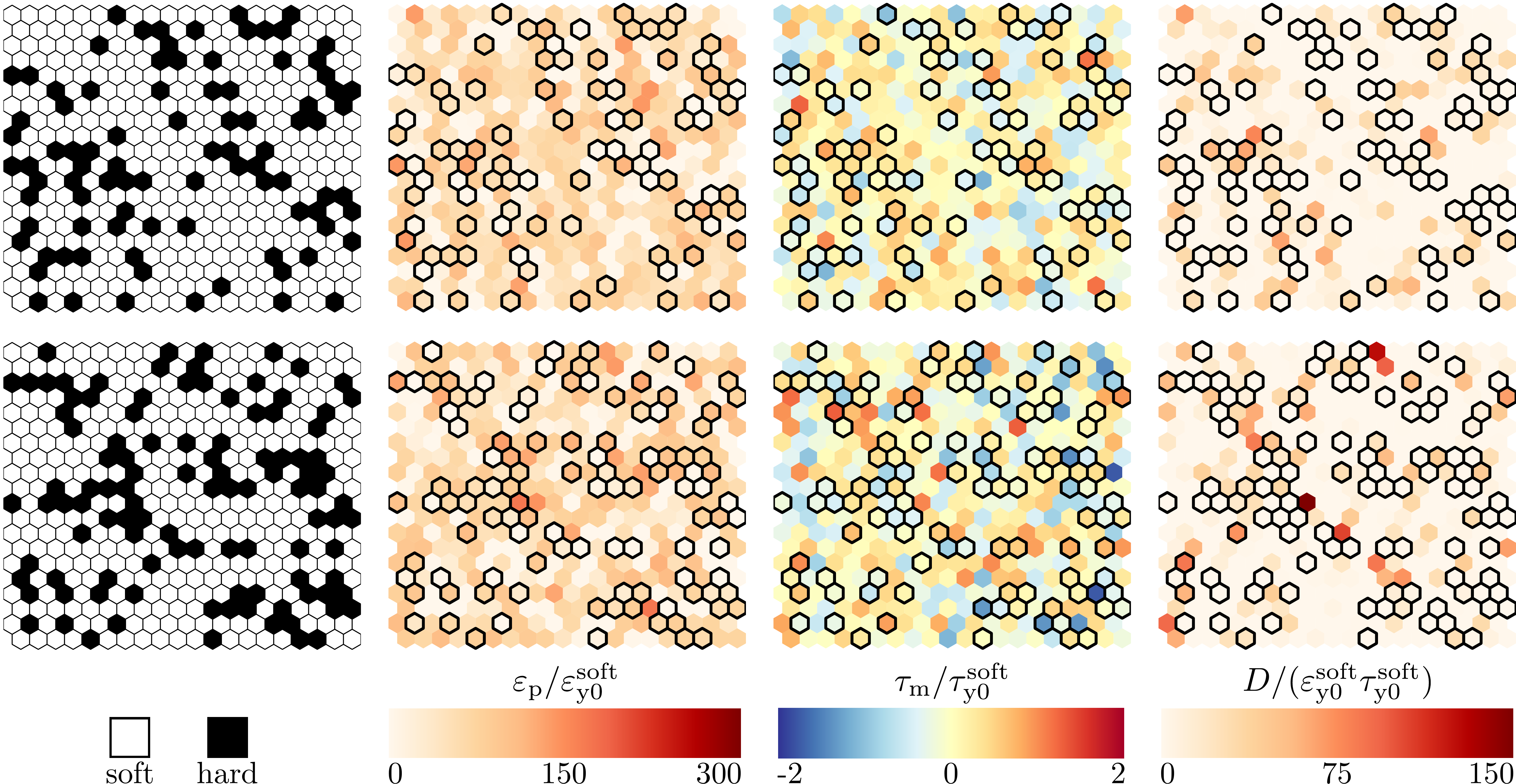}
  \caption{The individual response of the random cells displaying respectively the lowest (top) and highest damage (bottom) of the ensemble at the final increment of deformation. From left to right: the microstructure, the equivalent plastic strain $\varepsilon_\text{p}$, the hydrostatic stress $\tau_\text{m}$ and the damage $D$, normalized by the initial yield strain and stress ($\varepsilon_\text{y0}^\text{soft}$ and $\tau_\text{y0}^\text{soft}$) of the soft phase.}
  \label{fig:results}
\end{figure}

We consider the influence of topology in more detail by considering two individual cells at the two ends of the damage spectrum. Figure~\ref{fig:results} shows the topologies in which the damage is lowest (top) and highest (bottom) at the final increment of applied deformation. The figure shows from left to right: the topology, and the normalized equivalent plastic strain $\varepsilon_\text{p}$, hydrostatic stress $\tau_\text{m}$ and damage $D$. The principal tensile strain axis is horizontal. In these plots we observe that most of the plastic deformation is taken up by the soft phase, in particular close to hard particles. The largest hydrostatic tension is observed in the soft phase close hard particles. The resulting damage is therefore found mostly in the soft phase near hard particles, specifically where the hard particles are located adjacent to a soft element in the principal strain direction.

\begin{figure}[htp]
  \centering
  \begin{minipage}[t]{.31\textwidth}
  	\includegraphics[height=50mm]{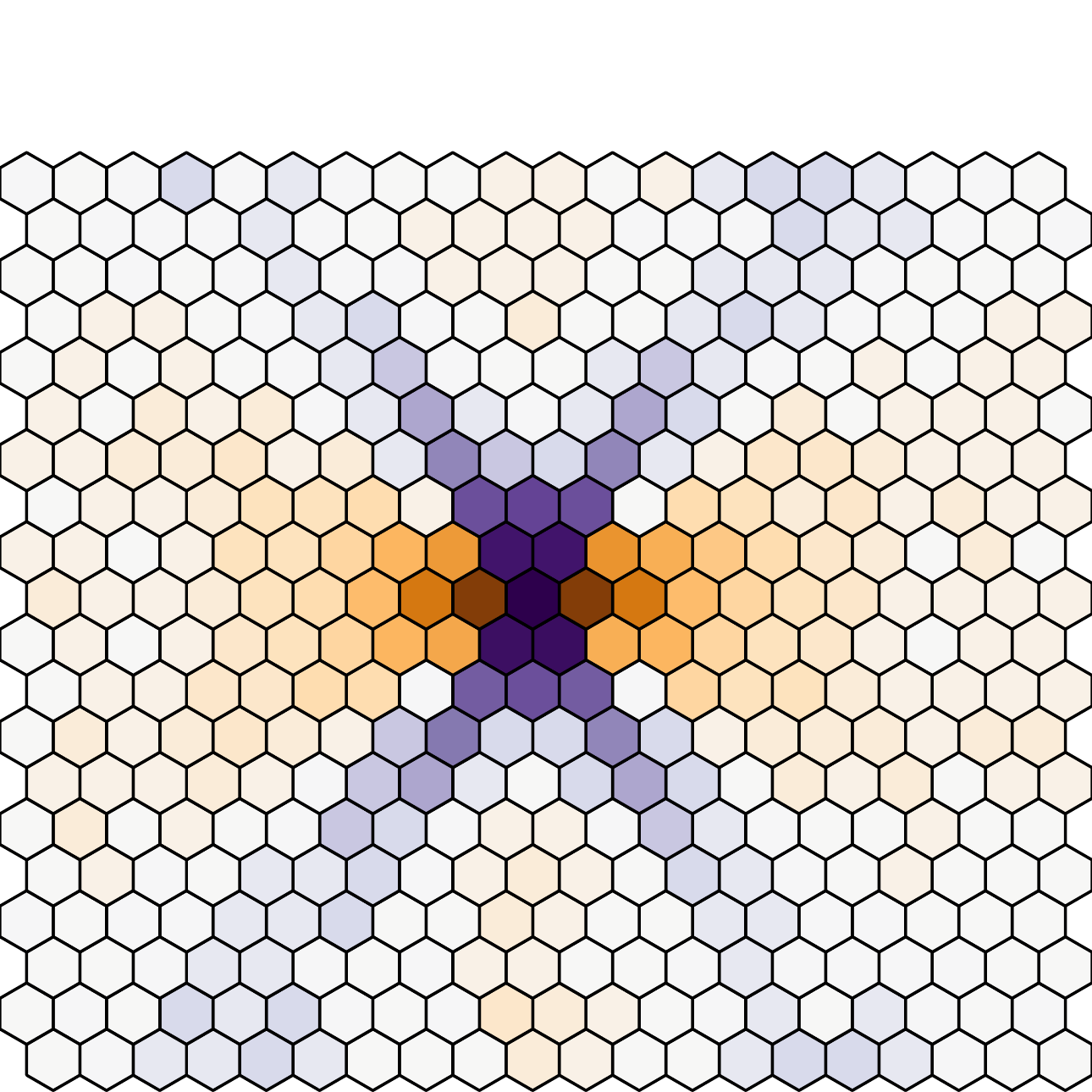}
  	\\ \footnotesize
  	(a)
  \end{minipage}
  \hspace{.05\textwidth}
  \begin{minipage}[t]{.31\textwidth}
  	\includegraphics[height=50mm]{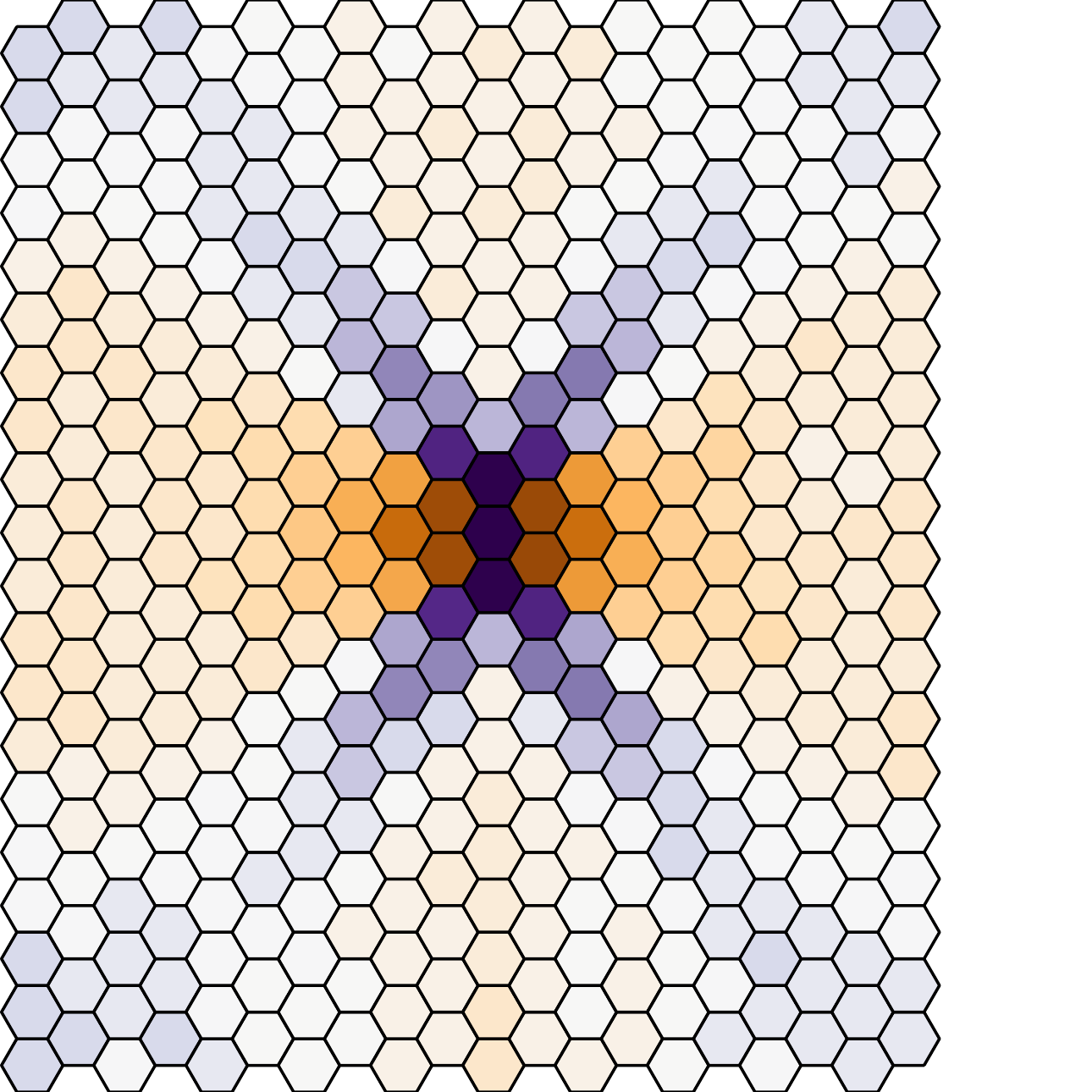}
  	\\ \footnotesize
  	(b)
  \end{minipage}
  \begin{minipage}[t]{.11\textwidth}
  	\includegraphics[height=50mm]{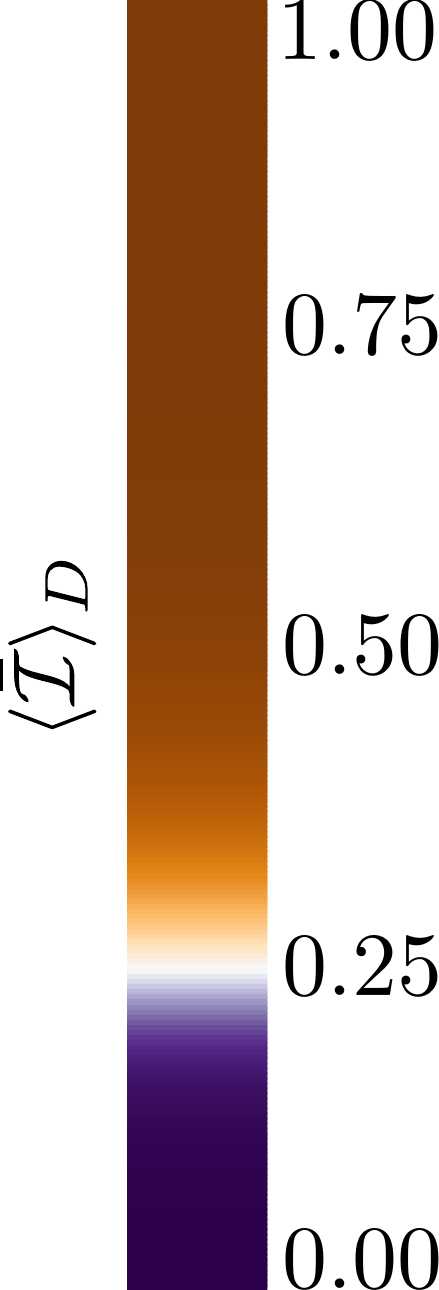}
  \end{minipage}
  \caption{The ensemble average indicator function $\langle \bar{\mathcal{I}}_D \rangle$ weighed by the damage, whereby the origin is arbitrarily chosen in the center. The neutral color of the colormap is chosen to coincide with the ensemble average hard phase volume fraction $\langle \varphi^\text{hard} \rangle$, and increases exponentially in both directions. The principal strain axis is in horizontal direction; the hexagon (and unit cell) orientation is rotated by $90^\circ$ between (a) and (b).}
  \label{fig:damage-stat}
\end{figure}

Based on the individual responses it is difficult to systematically determine the influence of the topology on the damage. We therefore consider all cells in the ensemble and calculate the probability of finding the hard phase around a damage ``hot-spot''. The method is detailed in \cite{DeGeus2015a}, and summarized here for completeness. We describe the distribution of phases using a so-called indicator function defined as follows
\begin{equation}
  \mathcal{I} (i,j) =
  \begin{cases}
    1 \quad &\text{if}~~(i,j) \in \text{hard}
    \\
    0 \quad &\text{if}~~(i,j) \in \text{soft}
  \end{cases}
\end{equation}
whereby $(i,j)$ corresponds to the ``matrix-position'' in the cell. The average distribution -- weighed by the damage in the element $(i,j)$ -- at a distance $(\Delta i, \Delta j)$ then reads
\begin{equation}
  \bar{\mathcal{I}}_D ( \Delta i , \Delta j ) = \frac{\sum_{i,j} D(i,j) \; \mathcal{I} ( i + \Delta i , j + \Delta j )}{\sum_{i,j} D(i,j)}
\end{equation}
whereby in the case of hexagons even and odd $i$ require a slightly different treatment. The damage averaged relative indicator function $\bar{\mathcal{I}}_D$ can be interpreted as the probability of finding the hard phase at a distance $(\Delta i, \Delta j)$ around highly damaged elements. Its ensemble average $\langle \bar{\mathcal{I}}_D \rangle$ is shown in Figure~\ref{fig:damage-stat}(a). $\langle \bar{\mathcal{I}}_D \rangle = 0$ corresponds to certainty of finding the soft phase and $\langle \bar{\mathcal{I}}_D \rangle = 1$ to the hard phase, around the hot-spot in the center of the diagram. From Figure~\ref{fig:damage-stat}(a) we observe that high damage consistently occurs in the soft phase, with on both sides regions of hard phase in the principal strain direction. In the opposite direction we find soft phase adjacent to the damage ``hot-spot'', and at angles close to $\pm 45$ degrees we observe an elevated probability of soft phase. Farther away, we observe some scatter around $\langle \bar{\mathcal{I}}_D \rangle = \langle \varphi^\text{hard} \rangle = 0.25$, the ensemble averaged hard phase volume fraction. The observed probability distribution is consistent with that found in \cite{DeGeus2015a} using square elements.

In the numerical results presented here -- as well as in \cite{DeGeus2015a} -- we observe close to the corners of the cell a slightly elevated probability of soft phase (i.e.\ a lower than average probability of hard phase). This effect is related to the assumed periodicity of the unit cells. Therefore, some caution is in order as we know from the literature (e.g.\ \cite{Coenen2012}) that the deformation may be sensitive to the periodicity -- in particular in the case of softening.

In the results presented here, this artifact is unlikely to be observed since the cells are non-square due to the nature of the hexagonal element. To further verify this, we rotate the cells, thereby changing the orientation. The result is shown in Figure~\ref{fig:damage-stat}(b). In comparing this to the earlier result in Figure~\ref{fig:damage-stat}(a) we observe only small differences, mostly related to the configuration of the individual elements. In particular observe that the region of elevated hard phase probability is slightly larger. The bands of elevated soft phase probability have more or less the same orientation, with small differences due to the local configuration of elements.

\section{Conclusions}
\label{sec:conclusion}

\begin{itemize}
\item The damage varies strongly with the topology of the microstructure. It appears to be uncorrelated to small variations of the volume fraction of individual cells.
\item By considering a large ensemble of random topologies we have obtained a probability distribution of hard and soft phase around a damage ``hot-spot''. It consists of hard phase  particles adjacent to a central soft element, in the principal strain direction, and bands of soft phase under $\pm 45$ degree angles to this direction.
\item We have confirmed the conclusions drawn in \cite{DeGeus2015a} using hexagonal instead of square elements.
\end{itemize}

\section*{Acknowledgments}

This research was carried out under project number M22.2.11424 in the framework of the Research Program of the Materials innovation institute M2i (\href{http://www.m2i.nl}{www.m2i.nl}).

J.~van Beeck and B.G.~Vossen are gratefully acknowledged for their assistance with the numerical implementation.


\bibliography{library}

\end{document}